\documentclass[aps,prl,twocolumn,groupedaddress,unsortedaddress,superscriptaddress]{revtex4-2}

\usepackage{amsmath} 
\usepackage{breqn}
\usepackage{amssymb}
\usepackage{siunitx}
\usepackage{graphicx}
\usepackage{hyperref} 

\makeatletter
\let\cat@comma@active\@empty
\makeatother

\begin{document} 

\title{Unambiguously Resolving the Potential Neutrino Magnetic Moment Signal at Large Liquid Scintillator Detectors} 

\author{Ziping Ye}
\affiliation{Tsung-Dao Lee Institute, Shanghai 200240, China}

\author{Feiyang Zhang}
\affiliation{INPAC and School of Physics and Astronomy, Shanghai Jiao Tong University, \\ 
Shanghai Laboratory for Particle Physics and Cosmology, \\ 
Key Laboratory for Particle Physics and Cosmology (MOE), Shanghai 200240, China}

\author{Donglian Xu}
\thanks{Corresponding author : \href{mailto:donglianxu@sjtu.edu.cn}{donglianxu@sjtu.edu.cn}}
\affiliation{Tsung-Dao Lee Institute, Shanghai 200240, China}
\affiliation{INPAC and School of Physics and Astronomy, Shanghai Jiao Tong University, \\ 
Shanghai Laboratory for Particle Physics and Cosmology, \\ 
Key Laboratory for Particle Physics and Cosmology (MOE), Shanghai 200240, China}

\author{Jianglai Liu}
\affiliation{INPAC and School of Physics and Astronomy, Shanghai Jiao Tong University, \\ 
Shanghai Laboratory for Particle Physics and Cosmology, \\ 
Key Laboratory for Particle Physics and Cosmology (MOE), Shanghai 200240, China}
\affiliation{Tsung-Dao Lee Institute, Shanghai 200240, China}

\begin{abstract}
    Non-vanishing electromagnetic properties of neutrinos have been predicted by many theories beyond the Standard Model, and an enhanced neutrino magnetic moment can have profound implications for fundamental physics. The XENON1T experiment recently detected an excess of electron recoil events in the 1-7 keV energy range, which can be compatible with solar neutrino magnetic moment interaction at a most probable value of $\mu_{\nu} = 2.1 \times 10^{-11} \mu_{\text{B}}$. However, tritium backgrounds or solar axion interaction in this energy window are equally plausible causes. Upcoming multi-tonne noble liquid detectors will test these scenarios more in depth, but will continue to face similar ambiguity. We report a unique capability of future large liquid scintillator detectors to help resolve the potential neutrino magnetic moment scenario. With $O$(100) kton$\cdot$year exposure of liquid scintillator to solar neutrinos, a sensitivity of $\mu_{\nu} < 10^{-11} \mu_{\text{B}}$ can be reached at an energy threshold greater than 40 keV, where no tritium or solar axion events but only neutrino magnetic moment signal is still present. 
\end{abstract} 


\maketitle

{\it Introduction} - The non-zero neutrino mass indicates that neutrinos have a tiny induced magnetic moment proportional to its mass, $\mu_{\nu} \approx 3.2 \times 10^{-19} (\, \frac{m_{\nu}}{1 eV} )\, \mu_B$ ($\mu_B = \frac{e \hbar}{2 m_e}$ is the Bohr magneton) which is so small that it is out of current experimental reach \cite{Marciano_Sanda_vMM_1977, Lee_Shrock_vMM_1977, Fujikawa_Shrock_vMM_1980}. However, many theories beyond the Standard Model predict that neutrinos can have much larger enhanced magnetic moments \cite{model_for_large_vMM_1987, model_for_large_vMM_1989, model_for_large_vMM_1998, Giunti_Studenikin_vMM_2009}. The enhanced magnetic moment allows neutrinos to interact via electromagnetic force and the effect can be large enough for it to be observed by the current and next generation of detectors, in contrast to only weak interactions of neutrinos in the Standard Model. This has raised tremendous interest both from the theoretical and experimental particle physics community. For a mysterious particle like the neutrino, which already holds a confirmed property beyond the Standard Model for being massive, measuring its magnetic moment is important both for further understanding the particle itself and for the search of new physics. By considering electroweak radiative corrections to neutrino mass generated by physics above the electroweak scale, it is shown in  \cite{Bell_Cirigliano_Ramsey_Vogel_Wise_2005_Dirac_vMM} that the current neutrino mass limit implies the magnetic moment of Dirac neutrinos can not exceed $\sim 10^{-14} \mu_B$, while for Majorana neutrinos there is no such strict limit on the transition magnetic moments \cite{Bell_Gorchtein_Ramsey_Vogel_Wang_2006_Majorana_vMM}. Therefore, if neutrinos are observed to have a magnetic moment larger than $\sim 10^{-14} \mu_B$, it will be a strong indication for Majorana particles. 

Experimental search for neutrino magnetic moment ($\nu$MM) began more than 40 years ago. Measuring the energy spectrum of electron recoils induced by neutrinos from nuclear reactors or the Sun has been the most sensitive probe to $\nu$MM in laboratories on Earth \cite{Reines_Gurr_Sobel_1976_veES}. The best sensitivity for reactor anti-neutrino $\nu$MM was reported by the GEMMA experiment in 2012, an upper limit of $\mu_{\nu} < 2.9 \times 10^{-11} \mu_B$ ($90\%$ C.L.) was derived for electron-flavor anti-neutrino based on null results \cite{GEMMA_Collaboration_vMM_2012}. The most sensitive search for solar neutrino $\nu$MM was reported by the Borexino experiment in 2017, which yielded an upper limit of $\mu_{\nu} < 2.8 \times 10^{-11} \mu_B$ ($90\%$ C.L.) for $\nu$MM of oscillation-mixed solar neutrinos, based on a null result \cite{Borexino_Collaboration_vMM_2017}. 
Astrophysical analyses with stellar cooling and supernovae have derived more stringent constraints for $\nu$MM than that measured in laboratories on Earth \cite{limit_vMM_SN1987A_1988, limit_vMM_solar_anti-neutrinos_2004, limit_vMM_stellar_cooling_2019}. These limits, however, involves complex assumptions and are heavily model-dependent. They often have to assume the existence of the yet to be confirmed right-handed neutrinos with masses in narrow ranges \cite{limit_vMM_models_2021}. Therefore, direct laboratory measurement is indispensable for obtaining decisive knowledge of $\nu$MM. 

\begin{figure}[!th]
    \centering
    \includegraphics[width=1.0\linewidth]{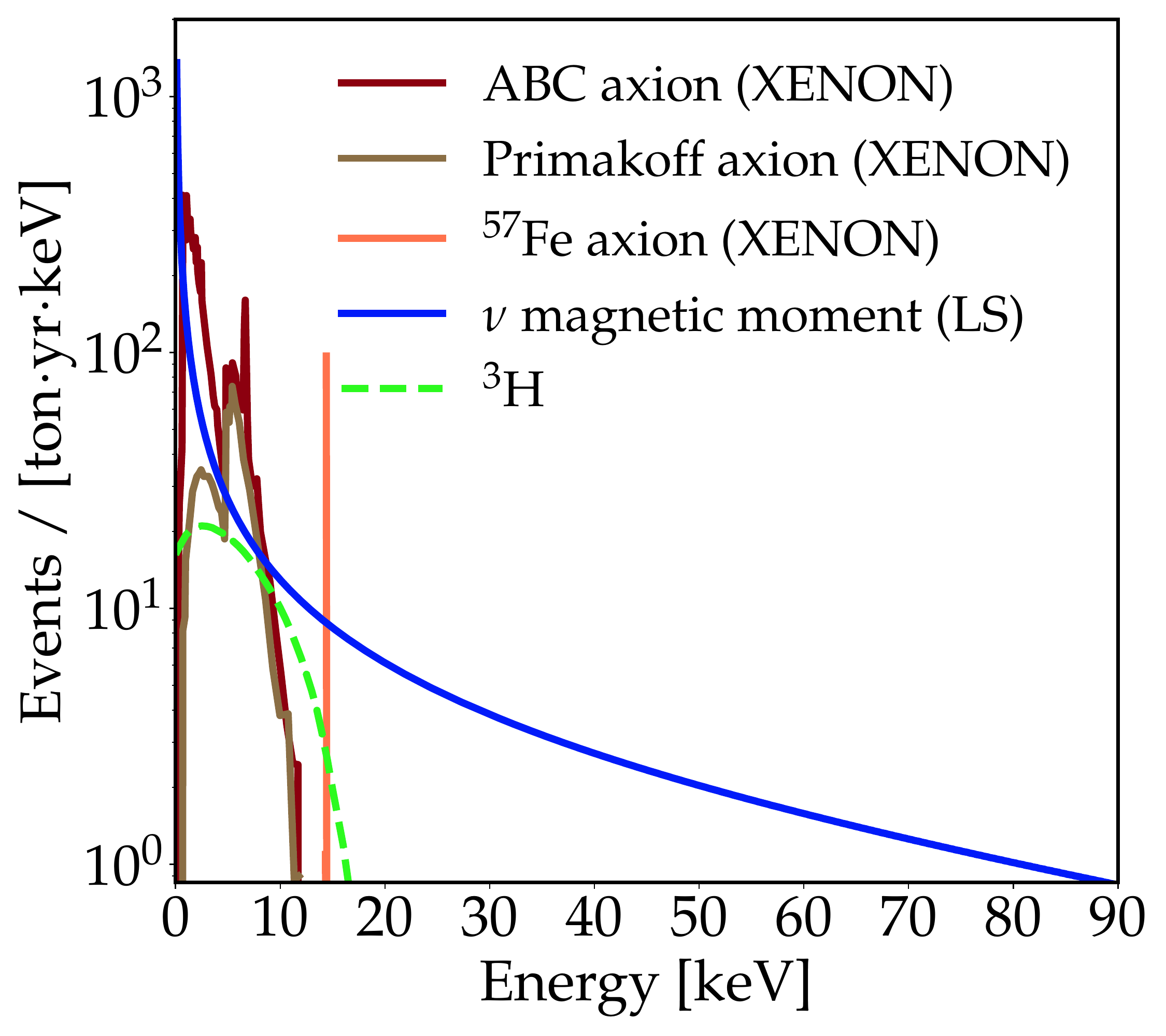}
    \caption{Comparison on signal strengths for (1) ABC solar axion model with a coupling constant $g_{ae} = 5 \times 10^{-12}$ (crimson), (2) $^{57}$Fe de-excitation axion model with a coupling constant $g_{an}^{eff} = 1 \times 10^{-6}$ (dirt), (3) Primakoff axion model with a coupling constant $g_{a{\gamma}} = 2 \times 10^{-10}$ (red) in liquid xenon detectors \cite{SolarAxion2013, SolarAxion1995, SolarAxion1986}; (4) solar neutrino magnetic moment at a value of $\mu_{\nu} = 2.1 \times 10^{-11} \mu_{\text{B}}$ (blue) in liquid scintillator detectors; and (5) the $\beta$ decay energy spectrum of tritium backgrounds at a level similar to the cosmogenic tritium isotope in JUNO \cite{JUNOYellowBook2016}. The number of electrons per ton in liquid xenon is about $73\%$ of that in linear alkylbenzene (LAB) based liquid scintillator.} 
    \label{Figure: xeon juno ls comparison}
\end{figure}

In 2020, the XENON1T experiment detected an excess of low energy electron recoil events in the 1-7 keV energy range, which is consistent with a $\nu$MM of $\mu_{\nu} \in (\, 1.4 , 2.9 )\, \times 10^{-11} \mu_B$ at $90\%$ C.L. \cite{XENON1T_Collaboration_excess_2020}. This excess, however, can also be explained by solar axions or backgrounds like tritium $\beta$-decays. While these hypotheses are equally compatible with the observed excess in the 1-7 keV energy range, the $\nu$MM scenario, if real, can produce signals that extend to energy regions well above 20 keV (see Fig.~\ref{Figure: xeon juno ls comparison}) and therefore can be cross validated unambiguously by large liquid scintillator (LS) detectors like JUNO \cite{JUNOYellowBook2016} and LENA \cite{Wurm:2011zn} with low detection thresholds. Although the upcoming multi-tonne xenon detectors, i.e. XENONnT \cite{Aprile_2020,xenon_dark_matter_detectors_vMM_2020}, LZ \cite{Akerib:2021pfd,Akerib:2021qbs} and PandaX-4T \cite{PandaX4T_dark_matter_detector_2019}, will be able to investigate this low energy excess more robustly, they continue to face the difficulty of ‘unexpected but possible’ backgrounds of $^3$H and $^{37}$Ar in this energy window \cite{xenon_dark_matter_detectors_vMM_2021}. Joint analyses between the forthcoming large liquid scintillator and liquid xenon detectors will help fully elucidate the potential signal of electromagnetic interaction of the solar neutrinos. 

The measurement of muon anomalous magnetic moment, $a_{\mu} \equiv ( g_{\mu} - 2 ) / 2 $, by the E821 experiment in 2006 at Brookhaven National Lab and the E989 experiment in 2021 at Fermilab revealed a combined 4.2$\sigma$ discrepancy between experiments and SM theory \cite{muon_g-2_BNL_2006, muon_g-2_Fermilab_2021}. Interestingly, it is suggested that new physics models which naturally generate large neutrino transition magnetic moment and small neutrino mass can also explain the muon g-2 anomaly, producing the right magnitude of signals in both XENON1T and muon g-2 \cite{muon_g-2_and_vMM_2021}. Direct and independent measurements of $\nu$MM thus become highly anticipated to shed light on these newly emerged puzzles.

%\section{SignalBackgrounds} 
{\it Neutrino magnetic moment signals and relevant backgrounds} - Solar neutrinos are detected via neutrino electron elastic scattering ($\nu$eES) in LS detectors. The energy spectrum of $\nu$eES events induced by solar neutrinos can be calculated by 

\begin{equation} \label{Equation: veES event rate}
    \frac{dN}{dT} = \int N_e \times \frac{d{\sigma(E_\nu | T)}}{dT} \times f(E_\nu) d{E_\nu} ,  
\end{equation}

where $T$ is the electron recoil energy, $N_e$ is the number of electrons in the LS target, $\frac{d{\sigma(E_\nu | T)}}{dT}$ is the differential cross section of $\nu$eES, and $f(E_\nu) d{E_\nu}$ is the solar neutrino flux in the energy interval $[ E_\nu , E_\nu + d{E_\nu} ]$. 

Besides the electron density of LS, the $\nu$eES event rate is determined by the $\nu$eES cross section and the solar neutrino flux. 

The $\nu$eES cross section predicted by the Standard Model (SM) is given by 

\begin{dmath} \label{Equation: veES cross section}
    \frac{d{\sigma^{SM}}}{dT} = \frac{G^2_F m_e}{2 \pi} [\, {(\, g_V + g_A )\,}^2 + (\, g^2_A - g^2_V )\, \frac{m_e T}{E^2_{\nu}} + {(\, g_V - g_A )\,}^2 {(\, 1 - \frac{T}{E_{\nu}} )\,}^2 ]\,   ,  
\end{dmath} 

where $G_F$ is the Fermi coupling constant, $m_e$ is the electron mass. For $\nu_e$, $g_V = 2 sin^2\theta_W + \frac{1}{2}$, $g_A = \frac{1}{2}$; for $\nu_x$ ($x = \mu$ or $\tau$), $g_V = 2 sin^2\theta_W - \frac{1}{2}$, $g_A = - \frac{1}{2}$ \cite{veES_cross_section_measurement_1993, veES_cross_section_measurement_2001, veES_cross_section_theory_2009, veES_cross_section_theory_2020}. 

If neutrinos have magnetic moment, they can interact via electromagnetic force, in addition to the weak force predicted by the SM. The $\nu$eES cross section will be the sum of the SM contribution and $\nu$MM contribution: 

\begin{equation}
    \frac{d{\sigma}}{dT} = \frac{d{\sigma^{SM}}}{dT} + \frac{d{\sigma^{\mu_\nu}}}{dT}   ,  
\end{equation} 

The $\nu$MM contribution is given by 

\begin{equation} \label{Equation: vMM veES cross section} 
    \frac{d{\sigma^{\mu_\nu}}}{dT} = \frac{\pi \alpha^2_{em}}{m^2_e} ( \frac{\mu_\nu}{\mu_B} )^2 \frac{1 - T/E_\nu}{T}   , 
\end{equation} 

where $\alpha_{em}$ is the fine structure constant, $\mu_\nu$ is the $\nu$MM, $\mu_B$ is the Bohr magneton \cite{vMM_1989, vMM_2005, vMM_review_2009}. It should be pointed out that adding radiative correction will result in about $2\%$ decrease of the SM cross section $\frac{d{\sigma^{SM}}}{dT}$, while $\nu$MM leads to an increase of the total cross section. In this sense, adding radiative correction will not affect the significance of the search for $\nu$MM, so in the following we will just take Equation \ref{Equation: veES cross section} as the SM $\nu$eES cross section, for simplicity. 

The solar neutrino fluxes have been measured by experiments \cite{Homestake_solar_neutrinos_1998, SNO_solar_neutrinos_2001, superK_solar_neutrinos_2001, Borexino_solar_pp_chain_neutrinos_2018, Borexino_solar_CNO_cycle_neutrinos_2020}, and their spectra are calculated by the Standard Solar Model (SSM) \cite{Standard_Solar_Model_1982, Standard_Solar_Model_1992, Standard_Solar_Model_2006, new_Standard_Solar_Model_2017}. We used the solar neutrino spectra from John Bahcall's homepage \cite{John_Bahcall_website_1} and the fluxes from high-metalicity SSM which is slightly favored by data \cite{new_Standard_Solar_Model_2017, Borexino_solar_pp_chain_neutrinos_2018, Borexino_solar_CNO_cycle_neutrinos_2020}. Since the search for $\nu$MM is to find potential deviation of the $\nu$eES energy spectrum from that predicted by the SM, we treat the solar neutrino flux as a single spectrum that is the sum of all the components (pp, pep, hep, $^7$Be, $^8$B from the pp-chain, and $^{13}$N, $^{15}$O, $^{17}$F from the CNO-cycle). The SM $\nu$eES cross sections for $\nu_e$ and $\nu_x$ ($x = \mu$ or $\tau$) are different, so the survival probability of $\nu_e$ due to flavor oscillation of solar neutrinos is taken into account for the calculation of $\nu$eES event rate \cite{Borexino_solar_pp_chain_neutrinos_2018}. 

A prominent feature of the $\nu$MM $\nu$eES signal is that it rises very fast as it goes to lower energies, due to the fact that $\nu$MM $\nu$eES cross section is inversely proportional to electron recoil energy ( $\frac{d{\sigma^{\mu_\nu}}}{dT} \propto 1/T$) and low energy solar neutrinos have a larger flux. Therefore, great advantage for probing $\nu$MM can be gained if the large LS detectors can lower its detection threshold. 

A reasonable estimate of the detector backgrounds is essential for the evaluation of large LS detectors' potential in the search of $\nu$MM. Besides the SM $\nu$eES events, the other relevant backgrounds include internal radioactivity, external radioactivity, and cosmogenic backgrounds. 

Internal radioactivity is mainly due to the isotopes: $^{14}$C, $^{238}$U, $^{232}$Th, $^{40}$K, $^{210}$Pb, $^{210}$Bi, $^{210}$Po and $^{85}$Kr \cite{JUNOYellowBook2016}. The inherent $^{14}$C is a major background for searching low energy signals in large LS detectors. This work assumes a $^{14}$C abundance of $^{14}$C/$^{12}$C = $1.0 \times 10^{-17}$; for comparison, the Borexino experiment reports an abundance of $^{14}$C/$^{12}$C = $0.27 \times 10^{-17}$ \cite{Borexino_C14_1998, Borexino_solar_pp_chain_C14_2014}. We also vary the $^{14}$C abundance by 5 times smaller or 5 times larger to evaluate its effect on the sensitivity. The pile-up of $^{14}$C backgrounds was taken into account in these evaluation. The rest of the radioactive isotopes are just impurities to the LS and can be reduced by purification. The contents of these impurities are referenced to the Table 13-11 in \cite{JUNOYellowBook2016}. Using the mature background suppression technique of pulse shape discrimination (PSD) in LS detectors, more than $99\%$ of $\alpha$ backgrounds can be removed, with $\sim 95\%$ efficiency of selecting $\beta$ and electron recoil events \cite{PSD_liquid_scintillator}. Because of this, only the $\beta$ and $\gamma$ backgrounds from $^{238}$U, $^{232}$Th and $^{210}$Pb decay chains are considered. 

External backgrounds are $\gamma$-rays emitted by the radioactive isotopes in the ambient environment. They can be shielded by veto detectors surrounding the central detector, and can be further reduced by fiducial cut in the central detector. By doing a proper fiducial cut, the external backgrounds can be suppressed to negligible \cite{JUNO_solar_B8_neutrinos_2020}. 

Cosmogenic backgrounds are induced by cosmic ray muons, they include cosmogenic neutrons and cosmogenic isotopes. Muons themselves produce events that deposit much larger energy than $\nu$MM $\nu$eES do, and they can be tagged by the veto detectors, hence they are not considered to be major backgrounds \cite{muon_tracking_JUNO_2018}. Cosmogenic neutrons will be captured by the protons in LS about 200 $\mu$s after production, releasing a 2.2 MeV $\gamma$, so they can be removed by cutting a few miliseconds data after the corresponding muon events. Cosmogenic isotopes are considered as backgrounds if they satisfy the following criteria: the decay energy of the isotopes fall into the region of the $\nu$MM $\nu$eES signals; the half-life of the radioactive isotopes are longer than the time of cut after muon events; the isotope production rate is larger than or comparable to the $\nu$MM $\nu$eES event rate. Cosmogenic isotopes that pass these criteria include: $^{3}$H, $^{6}$He, $^{7}$Be, $^{11}$C, and $^{12}$B. The production rates of these isotopes can be referred to the Table 13-9 of \cite{JUNOYellowBook2016}, if the LS detector is at deeper underground, the production rates will be smaller. For large LS detectors, usually the detection threshold is of $O$(100) keV, higher than the $^{3}$H $\beta$-decay end-point energy, so the $^{3}$H backgrounds can be ignored. 

The most significant electron recoil energy window for the search of $\nu$MM is in [100, 300] keV, with major contributions from $^{7}$Be and pp neutrinos. The major backgrounds in this energy window are due to $^{14}$C pile-up, SM $\nu$eES events, $^{210}$Po, and $^{210}$Bi.

{\it Sensitivity and discovery potential} - A toy Monte Carlo simulation has been established to study the detector response for the $\nu$MM signal and all the backgrounds in large LS detectors. For each event, the number of photo-electrons ($N_{pe}$) is obtained by Poisson random numbers: $N_{pe} = Poisson( \overline{N}_{pe} )$, with $\overline{N}_{pe}$ as the mean of the Poisson distribution. The mean $\overline{N}_{pe}$ for an event is a product of the event energy $E_{evt}$ and the photo-electron yield $Y_{pe}$. This work uses JUNO's detector design as the baseline to estimate the photo-electron yield, for it is under construction and will begin taking data soon, but the physics case sustains upon variation of detector designs for large LS detectors. The photo-electron yield is given by: $Y_{pe} = Y_{c1} \times f_{nonlin} \times f_{nonuni}$, where $Y_{c1} \approx 1300$ PE / MeV is the photo-electron yield for 1 MeV $\beta$ events at the detector center, $f_{nonlin}$ is the non-linear effect of the detector response for different energies, and $f_{nonuni}$ represents the non-uniform effect of the detector response at different locations in the detector \cite{JUNO_calibration_2020}. The dark noise of photo-sensors is added to each event also assuming Poisson distribution \cite{JUNO_20_inch_PMT}. 

We set up an analysis pipeline to study sensitivity and discovery potential with the toy Monte Carlo simulation data. The spectrum shape of $\nu$MM signal is highly distinguishable from that of all the backgrounds, so we can use spectrum fitting to search for the $\nu$MM signal. We use the method of ${\chi}^2$ fit, with the ${\chi}^2$ defined by: ${\chi}^2 = \sum_{n=1}^{N} \frac{(D_n - M_n)^2}{D_n}$, where $D$ represents the data and $D_n$ is the number of events in the $n^{th}$ bin of the spectrum, $M$ represents the model and $M_n$ is the expected number of events in the $n^{th}$ bin of the model, $N$ is the total number of bins. The best fit model is obtained by minimization of the ${\chi}^2$ \cite{chi_square_fit_1952}. 

Since the $\nu$MM has a higher signal rate at lower energy, lowering the energy threshold of the analysis can lead to better sensitivity. The $^{3}$H $\beta$-decay or potential solar axion events are below 20 keV, but by mixing with fluctuating dark noise of the photo-sensors, some of these events can get higher than 20 keV. Simulations show that, by setting the analysis threshold above 40 keV, contamination from $^{3}$H $\beta$-decay or potential solar axion events can be eliminated, and the systematic uncertainty of detector response at low energy can be reduced. A large LS detector with a low-threshold trigger system (down to $\sim$ 40 keV) is ideal to provide data for the search of $\nu$MM \cite{MM_trigger_1}. 

The $\nu$MM signals are mostly below $\sim$ 1300 keV, which divides the backgrounds into two categories: class-1 consists of $^{14}$C (and its pile-up), $^{210}$Pb chain ($^{210}$Pb, $^{210}$Bi, $^{210}$Po), $^{85}$Kr and $^{7}$Be, which are completely within the energy range of [0, 1300] keV where the $\nu$MM signals reside; class-2 consists of $^6$He, $^{11}$C, $^{12}$B, $^{40}$K, $^{232}$Th chain and $^{238}$U chain, which can have events with energy well above 1300 keV. So we can adopt a two-step fitting strategy: for the first step, fit the data above 1300 keV to constrain the backgrounds in class-2; then in the second step, fit the data below 1300 keV, with the class-2 backgrounds fixed and set free the rest. If fitting with a null model (backgrounds-only) yields ${\chi}_0^2$ and fitting with an alternative model (mixture of backgrounds and $\nu$MM signals) gives ${\chi}_1^2$, the detection significance can be given by $\sigma = \sqrt{{\chi}_0^2 - {\chi}_1^2}$.  

After accumulating an exposure of 127 kton$\cdot$year (corresponds to $\sim$ 10 years for JUNO with $\sim$ 12.7 kton fiducial mass or $\sim$ 3.3 years for LENA with $\sim$ 39 kton fiducial mass), the LS detectors' sensitivity to $\nu$MM can reach $0.7, 0.9, 1.1 \times 10^{-11} \mu_{B}$ ($90\%$ C.L.) for the $^{14}$C abundance of $^{14}$C/$^{12}$C = $0.2, 1.0, 5.0 \times 10^{-17}$, respectively. With a sensitivity better than the $\nu$MM range indicated by the XENON1T excess, future large LS detectors can resolve the $\nu$MM scenario. 

\begin{figure}[!th] 
    \centering
    \includegraphics[width=1.0\linewidth]{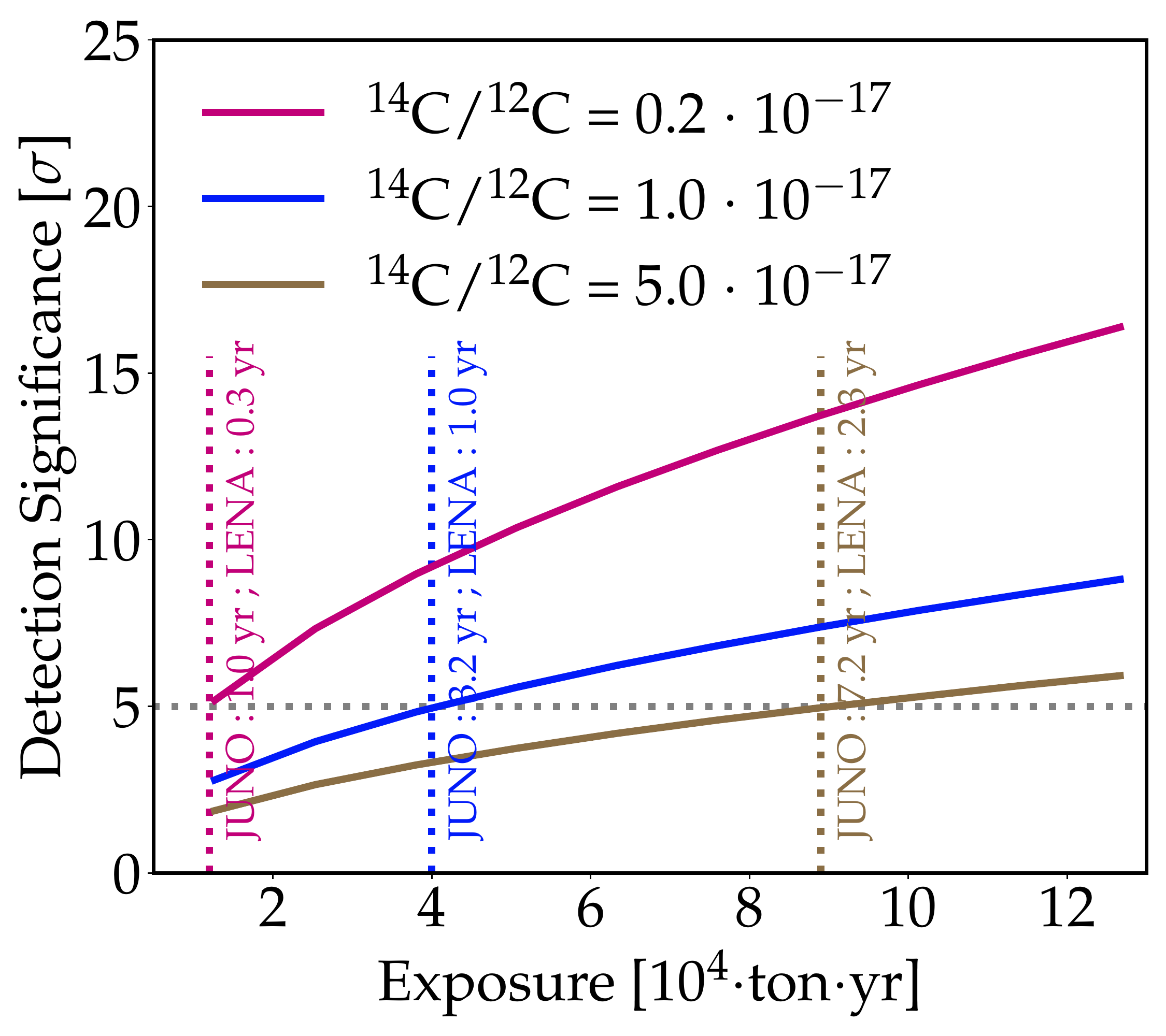}
    \caption{The detection significance, for a potential $\nu$MM signal with a most probable value compatible with the XENON1T excess ($\mu_{\nu} = 2.1 \times 10^{-11} \mu_{\text{B}}$), as a function of exposure for liquid scintillator (LS) detectors. Future large LS detectors like JUNO and LENA can reach these exposure. For both detectors, a fiducial cut that removes the outer layer of 2.5 m thickness is assumed to suppress the external backgrounds. For JUNO, the fiducial mass is 12.7 kton; for LENA, the fiducial mass is 39 kton. Both detectors possess the capability of unambiguously resolving the $\nu$MM scenario indicated by the XENON1T excess.}
    \label{Figure: Detection Significance JUNO Livetime} 
\end{figure} 

The discovery potential of large LS detectors to the $\nu$MM signal strength indicated by the XENON1T excess is shown in Figure \ref{Figure: Detection Significance JUNO Livetime} as a function of exposure. For the three values of $^{14}$C abundance mentioned above, a 5-$\sigma$ detection can be achieved for an exposure of 12.7, 40.9, 91.2 kton$\cdot$year, respectively. These exposures correspond to 1.0, 3.3, 7.2 years of JUNO livetime; and 0.3, 1.0, 2.3 years of LENA livetime. If JUNO or LENA find no $\nu$MM signal in the corresponding livetimes, the $\nu$MM explanation for the XENON1T excess will be rejected, then other physics scenarios are needed to be investigated.

{\it Summary and Discussion} - Exploration on the electromagnetic properties of neutrinos is important to test theories beyond the Standard Model. Neutrino magnetic moment, among other electromagnetic properties, is critical for understanding the nature of neutrinos and probing new physics. The next generation of LS detectors with extremely low threshold of $\sim$20 keV, made possible by their unprecedented large mass and modern photon detection technology, can conduct searches for neutrino magnetic moment in energy regions complementary to liquid xenon detectors. The intriguing excess of electron recoil events recently reported by XENON1T has stimulated heated debates on their origin. We have shown that large LS detectors, like the forthcoming JUNO and planned LENA, can unambiguously test the neutrino magnetic moment scenario. For an exposure of 127 kton$\cdot$year, future large LS detectors' sensitivity to $\nu$MM can reach $0.9 \times 10^{-11} \mu_{B}$ ($90\%$ C.L.) for an $^{14}$C abundance of $^{14}$C/$^{12}$C = $1.0 \times 10^{-17}$. With this scale of exposure, about $1.2 \times 10^{8}$, $2.2 \times 10^{7}$ and $2.3 \times 10^{5}$ solar neutrino events above 30, 300 and 2000 keV can be collected, which will be unprecedentedly large statistics and enable more accurate measurement of solar neutrino physics. 

Uncertainties of the study include that of solar neutrino fluxes, levels of $^{14}$C and other backgrounds, and the non-linearity of the detector response at low energies. The sensitivity to $\nu$MM is mainly due to $^{7}$Be and pp neutrinos which have only $\sim 0.6\%$ uncertainty in its flux, according to the Standard Solar Model, so the solar neutrino flux uncertainty is small. The spectral shape uncertainty of solar neutrinos and C-14 backgrounds are not severe problems. In the [100, 300] keV electron recoil energy window where $\nu$MM signal is most significant, the ratio of $\nu$MM signal to SM $\nu$eES backgrounds varies from $\sim 20\%$ to $5\%$. In the same energy window, it is not single $^{14}$C decay but $^{14}$C pile-up that is important. One can suppress $^{14}$C pile-up backgrounds by using pulse shape discrimination technique, so the $^{14}$C spectral shape uncertainty has minor effect on the detection significance, but its absolute level is more important. Since the $^{14}$C is embedded in the LS molecular structure, they cannot be reduced by purification as do for the other background isotopes. While LS extracted from deeper underground oil tend to contain lower $^{14}$C, it is nearly impossible to control the origin of oil in the industrial production line where LS is a by-product. Therefore, $^{14}$C level stays unknown till the LS detector is fully constructed and commissioned, hence it remains the largest uncertainty. As we have shown for three different levels of $^{14}$C backgrounds, from $^{14}$C/$^{12}$C = $0.2 \times 10^{-17}$ to 5 and 25 times higher, JUNO or LENA will be able to confirm or refute the neutrino magnetic moment scenario for the XENON1T excess in 7.2 or 2.3 years even for the worst case. The LS detector response in the low energy regime could be another main source of uncertainty, but it can be controlled by calibration using low energy radioactive sources. One promising low energy calibration source is the $^{57}$Co (122 keV) \cite{SNO_2016}, other sources like the very short-lived monoenergetic $^{83\text{m}}$Kr (41 keV) and $^{131\text{m}}$Xe (164 keV) used by liquid xenon experiments could potentially be employed by LS experiments as well \cite{Kastens_2009,Manalaysay_2010}. With most of these uncertainties under control and the variation of the major $^{14}$C background taken into account, the capability of large LS detectors in pinpointing the possible $\nu$MM interaction in XENON1T is proven to be robust.

\section*{Acknowledgement}
We thank the Strategic Priority Research Program of the Chinese Academy of Sciences, Grant No. XDA10010800, for the support to this work. DLX and ZPY are grateful to the Double First Class start-up fund (WF220442603) provided by Shanghai Jiao Tong University. DLX also thanks support from the CAS Center for Excellence in Particle Physics (CCEPP).

The authors thank Yifang Wang for helpful discussion on $^{14}$C background in liquid scintillator, and Tsutomu Yanagida for insightful conversations on neutrino magnetic moment. We also thank the JUNO Publication Committee, the two anonymous internal reviewers, and Jun Cao’s help to improve this paper.

\bibliography{references} 

\begin{thebibliography}{60}%
\makeatletter
\providecommand \@ifxundefined [1]{%
 \@ifx{#1\undefined}
}%
\providecommand \@ifnum [1]{%
 \ifnum #1\expandafter \@firstoftwo
 \else \expandafter \@secondoftwo
 \fi
}%
\providecommand \@ifx [1]{%
 \ifx #1\expandafter \@firstoftwo
 \else \expandafter \@secondoftwo
 \fi
}%
\providecommand \natexlab [1]{#1}%
\providecommand \enquote  [1]{``#1''}%
\providecommand \bibnamefont  [1]{#1}%
\providecommand \bibfnamefont [1]{#1}%
\providecommand \citenamefont [1]{#1}%
\providecommand \href@noop [0]{\@secondoftwo}%
\providecommand \href [0]{\begingroup \@sanitize@url \@href}%
\providecommand \@href[1]{\@@startlink{#1}\@@href}%
\providecommand \@@href[1]{\endgroup#1\@@endlink}%
\providecommand \@sanitize@url [0]{\catcode `\\12\catcode `\$12\catcode
  `\&12\catcode `\#12\catcode `\^12\catcode `\_12\catcode `\%12\relax}%
\providecommand \@@startlink[1]{}%
\providecommand \@@endlink[0]{}%
\providecommand \url  [0]{\begingroup\@sanitize@url \@url }%
\providecommand \@url [1]{\endgroup\@href {#1}{\urlprefix }}%
\providecommand \urlprefix  [0]{URL }%
\providecommand \Eprint [0]{\href }%
\providecommand \doibase [0]{https://doi.org/}%
\providecommand \selectlanguage [0]{\@gobble}%
\providecommand \bibinfo  [0]{\@secondoftwo}%
\providecommand \bibfield  [0]{\@secondoftwo}%
\providecommand \translation [1]{[#1]}%
\providecommand \BibitemOpen [0]{}%
\providecommand \bibitemStop [0]{}%
\providecommand \bibitemNoStop [0]{.\EOS\space}%
\providecommand \EOS [0]{\spacefactor3000\relax}%
\providecommand \BibitemShut  [1]{\csname bibitem#1\endcsname}%
\let\auto@bib@innerbib\@empty
%</preamble>
\bibitem [{\citenamefont {Marciano}\ and\ \citenamefont
  {Sanda}(1977)}]{Marciano_Sanda_vMM_1977}%
  \BibitemOpen
  \bibfield  {author} {\bibinfo {author} {\bibfnamefont {W.}~\bibnamefont
  {Marciano}}\ and\ \bibinfo {author} {\bibfnamefont {A.}~\bibnamefont
  {Sanda}},\ }\href@noop {} {\bibfield  {journal} {\bibinfo  {journal} {Physics
  Letters B}\ }\textbf {\bibinfo {volume} {67}},\ \bibinfo {pages} {303}
  (\bibinfo {year} {1977})}\BibitemShut {NoStop}%
\bibitem [{\citenamefont {Lee}\ and\ \citenamefont
  {Shrock}(1977)}]{Lee_Shrock_vMM_1977}%
  \BibitemOpen
  \bibfield  {author} {\bibinfo {author} {\bibfnamefont {B.~W.}\ \bibnamefont
  {Lee}}\ and\ \bibinfo {author} {\bibfnamefont {R.~E.}\ \bibnamefont
  {Shrock}},\ }\href@noop {} {\bibfield  {journal} {\bibinfo  {journal}
  {Physical Review D}\ }\textbf {\bibinfo {volume} {16}},\ \bibinfo {pages}
  {1444} (\bibinfo {year} {1977})}\BibitemShut {NoStop}%
\bibitem [{\citenamefont {Fujikawa}\ and\ \citenamefont
  {Shrock}(1980)}]{Fujikawa_Shrock_vMM_1980}%
  \BibitemOpen
  \bibfield  {author} {\bibinfo {author} {\bibfnamefont {K.}~\bibnamefont
  {Fujikawa}}\ and\ \bibinfo {author} {\bibfnamefont {R.~E.}\ \bibnamefont
  {Shrock}},\ }\href@noop {} {\bibfield  {journal} {\bibinfo  {journal}
  {Physical Review Letters}\ }\textbf {\bibinfo {volume} {45}},\ \bibinfo
  {pages} {963} (\bibinfo {year} {1980})}\BibitemShut {NoStop}%
\bibitem [{\citenamefont {Fukugita}\ and\ \citenamefont
  {Yanagida}(1987)}]{model_for_large_vMM_1987}%
  \BibitemOpen
  \bibfield  {author} {\bibinfo {author} {\bibfnamefont {M.}~\bibnamefont
  {Fukugita}}\ and\ \bibinfo {author} {\bibfnamefont {T.}~\bibnamefont
  {Yanagida}},\ }\href {https://doi.org/10.1103/PhysRevLett.58.1807} {\bibfield
   {journal} {\bibinfo  {journal} {Phys. Rev. Lett.}\ }\textbf {\bibinfo
  {volume} {58}},\ \bibinfo {pages} {1807} (\bibinfo {year}
  {1987})}\BibitemShut {NoStop}%
\bibitem [{\citenamefont {Babu}\ and\ \citenamefont
  {Mohapatra}(1989)}]{model_for_large_vMM_1989}%
  \BibitemOpen
  \bibfield  {author} {\bibinfo {author} {\bibfnamefont {K.~S.}\ \bibnamefont
  {Babu}}\ and\ \bibinfo {author} {\bibfnamefont {R.~N.}\ \bibnamefont
  {Mohapatra}},\ }\href {https://doi.org/10.1103/PhysRevLett.63.228} {\bibfield
   {journal} {\bibinfo  {journal} {Phys. Rev. Lett.}\ }\textbf {\bibinfo
  {volume} {63}},\ \bibinfo {pages} {228} (\bibinfo {year} {1989})}\BibitemShut
  {NoStop}%
\bibitem [{\citenamefont {Ma}(1998)}]{model_for_large_vMM_1998}%
  \BibitemOpen
  \bibfield  {author} {\bibinfo {author} {\bibfnamefont {E.}~\bibnamefont
  {Ma}},\ }\href {https://doi.org/10.1103/PhysRevLett.81.1171} {\bibfield
  {journal} {\bibinfo  {journal} {Phys. Rev. Lett.}\ }\textbf {\bibinfo
  {volume} {81}},\ \bibinfo {pages} {1171} (\bibinfo {year}
  {1998})}\BibitemShut {NoStop}%
\bibitem [{\citenamefont {Giunti}\ and\ \citenamefont
  {Studenikin}(2009{\natexlab{a}})}]{Giunti_Studenikin_vMM_2009}%
  \BibitemOpen
  \bibfield  {author} {\bibinfo {author} {\bibfnamefont {C.}~\bibnamefont
  {Giunti}}\ and\ \bibinfo {author} {\bibfnamefont {A.}~\bibnamefont
  {Studenikin}},\ }\href@noop {} {\bibfield  {journal} {\bibinfo  {journal}
  {Physics of Atomic Nuclei}\ }\textbf {\bibinfo {volume} {72}},\ \bibinfo
  {pages} {2089} (\bibinfo {year} {2009}{\natexlab{a}})}\BibitemShut {NoStop}%
\bibitem [{\citenamefont {Bell}\ \emph {et~al.}(2005)\citenamefont {Bell},
  \citenamefont {Cirigliano}, \citenamefont {Ramsey-Musolf}, \citenamefont
  {Vogel},\ and\ \citenamefont
  {Wise}}]{Bell_Cirigliano_Ramsey_Vogel_Wise_2005_Dirac_vMM}%
  \BibitemOpen
  \bibfield  {author} {\bibinfo {author} {\bibfnamefont {N.~F.}\ \bibnamefont
  {Bell}}, \bibinfo {author} {\bibfnamefont {V.}~\bibnamefont {Cirigliano}},
  \bibinfo {author} {\bibfnamefont {M.~J.}\ \bibnamefont {Ramsey-Musolf}},
  \bibinfo {author} {\bibfnamefont {P.}~\bibnamefont {Vogel}},\ and\ \bibinfo
  {author} {\bibfnamefont {M.~B.}\ \bibnamefont {Wise}},\ }\href@noop {}
  {\bibfield  {journal} {\bibinfo  {journal} {Physical review letters}\
  }\textbf {\bibinfo {volume} {95}},\ \bibinfo {pages} {151802} (\bibinfo
  {year} {2005})}\BibitemShut {NoStop}%
\bibitem [{\citenamefont {Bell}\ \emph {et~al.}(2006)\citenamefont {Bell},
  \citenamefont {Gorchtein}, \citenamefont {Ramsey-Musolf}, \citenamefont
  {Vogel},\ and\ \citenamefont
  {Wang}}]{Bell_Gorchtein_Ramsey_Vogel_Wang_2006_Majorana_vMM}%
  \BibitemOpen
  \bibfield  {author} {\bibinfo {author} {\bibfnamefont {N.~F.}\ \bibnamefont
  {Bell}}, \bibinfo {author} {\bibfnamefont {M.}~\bibnamefont {Gorchtein}},
  \bibinfo {author} {\bibfnamefont {M.~J.}\ \bibnamefont {Ramsey-Musolf}},
  \bibinfo {author} {\bibfnamefont {P.}~\bibnamefont {Vogel}},\ and\ \bibinfo
  {author} {\bibfnamefont {P.}~\bibnamefont {Wang}},\ }\href@noop {} {\bibfield
   {journal} {\bibinfo  {journal} {Physics Letters B}\ }\textbf {\bibinfo
  {volume} {642}},\ \bibinfo {pages} {377} (\bibinfo {year}
  {2006})}\BibitemShut {NoStop}%
\bibitem [{\citenamefont {Reines}\ \emph {et~al.}(1976)\citenamefont {Reines},
  \citenamefont {Gurr},\ and\ \citenamefont
  {Sobel}}]{Reines_Gurr_Sobel_1976_veES}%
  \BibitemOpen
  \bibfield  {author} {\bibinfo {author} {\bibfnamefont {F.}~\bibnamefont
  {Reines}}, \bibinfo {author} {\bibfnamefont {H.}~\bibnamefont {Gurr}},\ and\
  \bibinfo {author} {\bibfnamefont {H.}~\bibnamefont {Sobel}},\ }\href@noop {}
  {\bibfield  {journal} {\bibinfo  {journal} {Physical Review Letters}\
  }\textbf {\bibinfo {volume} {37}},\ \bibinfo {pages} {315} (\bibinfo {year}
  {1976})}\BibitemShut {NoStop}%
\bibitem [{\citenamefont {Beda}\ \emph {et~al.}(2012)\citenamefont {Beda},
  \citenamefont {Brudanin}, \citenamefont {Egorov}, \citenamefont {Medvedev},
  \citenamefont {Pogosov}, \citenamefont {Shirchenko},\ and\ \citenamefont
  {Starostin}}]{GEMMA_Collaboration_vMM_2012}%
  \BibitemOpen
  \bibfield  {author} {\bibinfo {author} {\bibfnamefont {A.}~\bibnamefont
  {Beda}}, \bibinfo {author} {\bibfnamefont {V.}~\bibnamefont {Brudanin}},
  \bibinfo {author} {\bibfnamefont {V.}~\bibnamefont {Egorov}}, \bibinfo
  {author} {\bibfnamefont {D.}~\bibnamefont {Medvedev}}, \bibinfo {author}
  {\bibfnamefont {V.}~\bibnamefont {Pogosov}}, \bibinfo {author} {\bibfnamefont
  {M.}~\bibnamefont {Shirchenko}},\ and\ \bibinfo {author} {\bibfnamefont
  {A.}~\bibnamefont {Starostin}},\ }\href@noop {} {\bibfield  {journal}
  {\bibinfo  {journal} {Advances in High Energy Physics}\ }\textbf {\bibinfo
  {volume} {2012}},\ \bibinfo {pages} {12} (\bibinfo {year}
  {2012})}\BibitemShut {NoStop}%
\bibitem [{\citenamefont {Agostini}\ \emph {et~al.}(2017)\citenamefont
  {Agostini}, \citenamefont {Altenm{\"u}ller}, \citenamefont {Appel},
  \citenamefont {Atroshchenko}, \citenamefont {Bagdasarian} \emph
  {et~al.}}]{Borexino_Collaboration_vMM_2017}%
  \BibitemOpen
  \bibfield  {author} {\bibinfo {author} {\bibfnamefont {M.}~\bibnamefont
  {Agostini}}, \bibinfo {author} {\bibfnamefont {K.}~\bibnamefont
  {Altenm{\"u}ller}}, \bibinfo {author} {\bibfnamefont {S.}~\bibnamefont
  {Appel}}, \bibinfo {author} {\bibfnamefont {V.}~\bibnamefont {Atroshchenko}},
  \bibinfo {author} {\bibfnamefont {Z.}~\bibnamefont {Bagdasarian}}, \emph
  {et~al.},\ }\href@noop {} {\bibfield  {journal} {\bibinfo  {journal}
  {Physical Review D}\ }\textbf {\bibinfo {volume} {96}},\ \bibinfo {pages}
  {091103} (\bibinfo {year} {2017})}\BibitemShut {NoStop}%
\bibitem [{\citenamefont {Lattimer}\ and\ \citenamefont
  {Cooperstein}(1988)}]{limit_vMM_SN1987A_1988}%
  \BibitemOpen
  \bibfield  {author} {\bibinfo {author} {\bibfnamefont {J.~M.}\ \bibnamefont
  {Lattimer}}\ and\ \bibinfo {author} {\bibfnamefont {J.}~\bibnamefont
  {Cooperstein}},\ }\href@noop {} {\bibfield  {journal} {\bibinfo  {journal}
  {Physical review letters}\ }\textbf {\bibinfo {volume} {61}},\ \bibinfo
  {pages} {23} (\bibinfo {year} {1988})}\BibitemShut {NoStop}%
\bibitem [{\citenamefont {Miranda}\ \emph {et~al.}(2004)\citenamefont
  {Miranda}, \citenamefont {Rashba}, \citenamefont {Rez},\ and\ \citenamefont
  {Valle}}]{limit_vMM_solar_anti-neutrinos_2004}%
  \BibitemOpen
  \bibfield  {author} {\bibinfo {author} {\bibfnamefont {O.}~\bibnamefont
  {Miranda}}, \bibinfo {author} {\bibfnamefont {T.~I.}\ \bibnamefont {Rashba}},
  \bibinfo {author} {\bibfnamefont {A.}~\bibnamefont {Rez}},\ and\ \bibinfo
  {author} {\bibfnamefont {J.}~\bibnamefont {Valle}},\ }\href@noop {}
  {\bibfield  {journal} {\bibinfo  {journal} {Physical review letters}\
  }\textbf {\bibinfo {volume} {93}},\ \bibinfo {pages} {051304} (\bibinfo
  {year} {2004})}\BibitemShut {NoStop}%
\bibitem [{\citenamefont {D{\'\i}az}\ \emph {et~al.}(2019)\citenamefont
  {D{\'\i}az}, \citenamefont {Schr{\"o}der}, \citenamefont {Zuber},
  \citenamefont {Jack},\ and\ \citenamefont
  {Barrios}}]{limit_vMM_stellar_cooling_2019}%
  \BibitemOpen
  \bibfield  {author} {\bibinfo {author} {\bibfnamefont {S.~A.}\ \bibnamefont
  {D{\'\i}az}}, \bibinfo {author} {\bibfnamefont {K.-P.}\ \bibnamefont
  {Schr{\"o}der}}, \bibinfo {author} {\bibfnamefont {K.}~\bibnamefont {Zuber}},
  \bibinfo {author} {\bibfnamefont {D.}~\bibnamefont {Jack}},\ and\ \bibinfo
  {author} {\bibfnamefont {E.~E.~B.}\ \bibnamefont {Barrios}},\ }\href@noop {}
  {\bibfield  {journal} {\bibinfo  {journal} {arXiv preprint arXiv:1910.10568}\
  } (\bibinfo {year} {2019})}\BibitemShut {NoStop}%
\bibitem [{\citenamefont {Brdar}\ \emph {et~al.}(2021)\citenamefont {Brdar},
  \citenamefont {Greljo}, \citenamefont {Kopp},\ and\ \citenamefont
  {Opferkuch}}]{limit_vMM_models_2021}%
  \BibitemOpen
  \bibfield  {author} {\bibinfo {author} {\bibfnamefont {V.}~\bibnamefont
  {Brdar}}, \bibinfo {author} {\bibfnamefont {A.}~\bibnamefont {Greljo}},
  \bibinfo {author} {\bibfnamefont {J.}~\bibnamefont {Kopp}},\ and\ \bibinfo
  {author} {\bibfnamefont {T.}~\bibnamefont {Opferkuch}},\ }\href@noop {}
  {\bibfield  {journal} {\bibinfo  {journal} {Journal of Cosmology and
  Astroparticle Physics}\ }\textbf {\bibinfo {volume} {2021}}\bibinfo  {number}
  { (01)},\ \bibinfo {pages} {039}}\BibitemShut {NoStop}%
\bibitem [{\citenamefont {Aprile}\ \emph
  {et~al.}(2020{\natexlab{a}})\citenamefont {Aprile}, \citenamefont {Aalbers},
  \citenamefont {Agostini}, \citenamefont {Alfonsi}, \citenamefont {Althueser}
  \emph {et~al.}}]{XENON1T_Collaboration_excess_2020}%
  \BibitemOpen
\bibfield  {number} {  }\bibfield  {author} {\bibinfo {author} {\bibfnamefont
  {E.}~\bibnamefont {Aprile}}, \bibinfo {author} {\bibfnamefont
  {J.}~\bibnamefont {Aalbers}}, \bibinfo {author} {\bibfnamefont
  {F.}~\bibnamefont {Agostini}}, \bibinfo {author} {\bibfnamefont
  {M.}~\bibnamefont {Alfonsi}}, \bibinfo {author} {\bibfnamefont
  {L.}~\bibnamefont {Althueser}}, \emph {et~al.},\ }\href@noop {} {\bibfield
  {journal} {\bibinfo  {journal} {Physical Review D}\ }\textbf {\bibinfo
  {volume} {102}},\ \bibinfo {pages} {072004} (\bibinfo {year}
  {2020}{\natexlab{a}})}\BibitemShut {NoStop}%
\bibitem [{\citenamefont {An}\ \emph {et~al.}(2016)\citenamefont {An},
  \citenamefont {An}, \citenamefont {An}, \citenamefont {Antonelli},
  \citenamefont {Baussan}, \citenamefont {Beacom} \emph
  {et~al.}}]{JUNOYellowBook2016}%
  \BibitemOpen
  \bibfield  {author} {\bibinfo {author} {\bibfnamefont {F.}~\bibnamefont
  {An}}, \bibinfo {author} {\bibfnamefont {G.}~\bibnamefont {An}}, \bibinfo
  {author} {\bibfnamefont {Q.}~\bibnamefont {An}}, \bibinfo {author}
  {\bibfnamefont {V.}~\bibnamefont {Antonelli}}, \bibinfo {author}
  {\bibfnamefont {E.}~\bibnamefont {Baussan}}, \bibinfo {author} {\bibfnamefont
  {J.}~\bibnamefont {Beacom}}, \emph {et~al.},\ }\href@noop {} {\bibfield
  {journal} {\bibinfo  {journal} {Journal of Physics G: Nuclear and Particle
  Physics}\ }\textbf {\bibinfo {volume} {43}},\ \bibinfo {pages} {030401}
  (\bibinfo {year} {2016})}\BibitemShut {NoStop}%
\bibitem [{\citenamefont {Wurm}\ \emph {et~al.}(2012)\citenamefont {Wurm} \emph
  {et~al.}}]{Wurm:2011zn}%
  \BibitemOpen
  \bibfield  {author} {\bibinfo {author} {\bibfnamefont {M.}~\bibnamefont
  {Wurm}} \emph {et~al.} (\bibinfo {collaboration} {LENA}),\ }\href
  {https://doi.org/10.1016/j.astropartphys.2012.02.011} {\bibfield  {journal}
  {\bibinfo  {journal} {Astropart. Phys.}\ }\textbf {\bibinfo {volume} {35}},\
  \bibinfo {pages} {685} (\bibinfo {year} {2012})},\ \Eprint
  {https://arxiv.org/abs/1104.5620} {arXiv:1104.5620 [astro-ph.IM]}
  \BibitemShut {NoStop}%
\bibitem [{\citenamefont {Aprile}\ \emph
  {et~al.}(2020{\natexlab{b}})\citenamefont {Aprile}, \citenamefont {Aalbers},
  \citenamefont {Agostini}, \citenamefont {Alfonsi}, \citenamefont {Althueser},
  \citenamefont {Amaro}, \citenamefont {Antochi}, \citenamefont {Angelino},
  \citenamefont {Angevaare}, \citenamefont {Arneodo},\ and\ \citenamefont
  {et~al.}}]{Aprile_2020}%
  \BibitemOpen
  \bibfield  {author} {\bibinfo {author} {\bibfnamefont {E.}~\bibnamefont
  {Aprile}}, \bibinfo {author} {\bibfnamefont {J.}~\bibnamefont {Aalbers}},
  \bibinfo {author} {\bibfnamefont {F.}~\bibnamefont {Agostini}}, \bibinfo
  {author} {\bibfnamefont {M.}~\bibnamefont {Alfonsi}}, \bibinfo {author}
  {\bibfnamefont {L.}~\bibnamefont {Althueser}}, \bibinfo {author}
  {\bibfnamefont {F.}~\bibnamefont {Amaro}}, \bibinfo {author} {\bibfnamefont
  {V.}~\bibnamefont {Antochi}}, \bibinfo {author} {\bibfnamefont
  {E.}~\bibnamefont {Angelino}}, \bibinfo {author} {\bibfnamefont
  {J.}~\bibnamefont {Angevaare}}, \bibinfo {author} {\bibfnamefont
  {F.}~\bibnamefont {Arneodo}},\ and\ \bibinfo {author} {\bibnamefont
  {et~al.}},\ }\href {https://doi.org/10.1088/1475-7516/2020/11/031} {\bibfield
   {journal} {\bibinfo  {journal} {Journal of Cosmology and Astroparticle
  Physics}\ }\textbf {\bibinfo {volume} {2020}}\bibinfo  {number} { (11)},\
  \bibinfo {pages} {031–031}}\BibitemShut {NoStop}%
\bibitem [{\citenamefont {Aristizabal~Sierra}\ \emph
  {et~al.}(2020)\citenamefont {Aristizabal~Sierra}, \citenamefont {Branada},
  \citenamefont {Miranda},\ and\ \citenamefont
  {Sanchez~Garcia}}]{xenon_dark_matter_detectors_vMM_2020}%
  \BibitemOpen
\bibfield  {number} {  }\bibfield  {author} {\bibinfo {author} {\bibfnamefont
  {D.}~\bibnamefont {Aristizabal~Sierra}}, \bibinfo {author} {\bibfnamefont
  {R.}~\bibnamefont {Branada}}, \bibinfo {author} {\bibfnamefont {O.~G.}\
  \bibnamefont {Miranda}},\ and\ \bibinfo {author} {\bibfnamefont
  {G.}~\bibnamefont {Sanchez~Garcia}},\ }\href
  {https://doi.org/10.1007/JHEP12(2020)178} {\bibfield  {journal} {\bibinfo
  {journal} {JHEP}\ }\textbf {\bibinfo {volume} {12}},\ \bibinfo {pages}
  {178}},\ \Eprint {https://arxiv.org/abs/2008.05080} {arXiv:2008.05080
  [hep-ph]} \BibitemShut {NoStop}%
\bibitem [{\citenamefont {Akerib}\ \emph
  {et~al.}(2021{\natexlab{a}})\citenamefont {Akerib} \emph
  {et~al.}}]{Akerib:2021pfd}%
  \BibitemOpen
  \bibfield  {author} {\bibinfo {author} {\bibfnamefont {D.~S.}\ \bibnamefont
  {Akerib}} \emph {et~al.},\ }\href@noop {} {\  (\bibinfo {year}
  {2021}{\natexlab{a}})},\ \Eprint {https://arxiv.org/abs/2101.08753}
  {arXiv:2101.08753 [astro-ph.IM]} \BibitemShut {NoStop}%
\bibitem [{\citenamefont {Akerib}\ \emph
  {et~al.}(2021{\natexlab{b}})\citenamefont {Akerib} \emph
  {et~al.}}]{Akerib:2021qbs}%
  \BibitemOpen
  \bibfield  {author} {\bibinfo {author} {\bibfnamefont {D.~S.}\ \bibnamefont
  {Akerib}} \emph {et~al.} (\bibinfo {collaboration} {LZ}),\ }\href@noop {} {\
  (\bibinfo {year} {2021}{\natexlab{b}})},\ \Eprint
  {https://arxiv.org/abs/2102.11740} {arXiv:2102.11740 [hep-ex]} \BibitemShut
  {NoStop}%
\bibitem [{\citenamefont {Zhang}\ \emph {et~al.}(2019)\citenamefont {Zhang}
  \emph {et~al.}}]{PandaX4T_dark_matter_detector_2019}%
  \BibitemOpen
  \bibfield  {author} {\bibinfo {author} {\bibfnamefont {H.}~\bibnamefont
  {Zhang}} \emph {et~al.} (\bibinfo {collaboration} {PandaX}),\ }\href
  {https://doi.org/10.1007/s11433-018-9259-0} {\bibfield  {journal} {\bibinfo
  {journal} {Sci. China Phys. Mech. Astron.}\ }\textbf {\bibinfo {volume}
  {62}},\ \bibinfo {pages} {31011} (\bibinfo {year} {2019})},\ \Eprint
  {https://arxiv.org/abs/1806.02229} {arXiv:1806.02229 [physics.ins-det]}
  \BibitemShut {NoStop}%
\bibitem [{\citenamefont {Ni}\ \emph {et~al.}(2021)\citenamefont {Ni},
  \citenamefont {Qi}, \citenamefont {Shockley},\ and\ \citenamefont
  {Wei}}]{xenon_dark_matter_detectors_vMM_2021}%
  \BibitemOpen
  \bibfield  {author} {\bibinfo {author} {\bibfnamefont {K.}~\bibnamefont
  {Ni}}, \bibinfo {author} {\bibfnamefont {J.}~\bibnamefont {Qi}}, \bibinfo
  {author} {\bibfnamefont {E.}~\bibnamefont {Shockley}},\ and\ \bibinfo
  {author} {\bibfnamefont {Y.}~\bibnamefont {Wei}},\ }\bibfield  {journal}
  {\bibinfo  {journal} {Universe}\ }\textbf {\bibinfo {volume} {7}},\ \href
  {https://doi.org/10.3390/universe7030054} {10.3390/universe7030054} (\bibinfo
  {year} {2021})\BibitemShut {NoStop}%
\bibitem [{\citenamefont {Muon~g 2~Collaboration: G.
  W.~Bennett}(2006)}]{muon_g-2_BNL_2006}%
  \BibitemOpen
  \bibfield  {author} {\bibinfo {author} {\bibfnamefont {e.~a.}\ \bibnamefont
  {Muon~g 2~Collaboration: G. W.~Bennett}} (\bibinfo {collaboration} {Muon g-2
  Collaboration}),\ }\href {https://doi.org/10.1103/PhysRevD.73.072003}
  {\bibfield  {journal} {\bibinfo  {journal} {Phys. Rev. D}\ }\textbf {\bibinfo
  {volume} {73}},\ \bibinfo {pages} {072003} (\bibinfo {year}
  {2006})}\BibitemShut {NoStop}%
\bibitem [{\citenamefont {g-2 Collaboration: B. Abi~et
  al.}(2021)}]{muon_g-2_Fermilab_2021}%
  \BibitemOpen
  \bibfield  {author} {\bibinfo {author} {\bibfnamefont {M.}~\bibnamefont {g-2
  Collaboration: B. Abi~et al.}} (\bibinfo {collaboration} {Muon
  $g\ensuremath{-}2$ Collaboration}),\ }\href
  {https://doi.org/10.1103/PhysRevLett.126.141801} {\bibfield  {journal}
  {\bibinfo  {journal} {Phys. Rev. Lett.}\ }\textbf {\bibinfo {volume} {126}},\
  \bibinfo {pages} {141801} (\bibinfo {year} {2021})}\BibitemShut {NoStop}%
\bibitem [{\citenamefont {Babu}\ \emph {et~al.}(2021)\citenamefont {Babu},
  \citenamefont {Jana}, \citenamefont {Lindner},\ and\ \citenamefont
  {K}}]{muon_g-2_and_vMM_2021}%
  \BibitemOpen
  \bibfield  {author} {\bibinfo {author} {\bibfnamefont {K.~S.}\ \bibnamefont
  {Babu}}, \bibinfo {author} {\bibfnamefont {S.}~\bibnamefont {Jana}}, \bibinfo
  {author} {\bibfnamefont {M.}~\bibnamefont {Lindner}},\ and\ \bibinfo {author}
  {\bibfnamefont {V.~P.}\ \bibnamefont {K}},\ }\href@noop {} {\bibinfo {title}
  {Muon ${g-2}$ anomaly and neutrino magnetic moments}} (\bibinfo {year}
  {2021}),\ \Eprint {https://arxiv.org/abs/2104.03291} {arXiv:2104.03291
  [hep-ph]} \BibitemShut {NoStop}%
\bibitem [{\citenamefont {Redondo}(2013)}]{SolarAxion2013}%
  \BibitemOpen
  \bibfield  {author} {\bibinfo {author} {\bibfnamefont {J.}~\bibnamefont
  {Redondo}},\ }\href@noop {} {\bibfield  {journal} {\bibinfo  {journal}
  {Journal of Cosmology and Astroparticle Physics}\ }\textbf {\bibinfo {volume}
  {2013}}\bibinfo  {number} { (12)},\ \bibinfo {pages} {008}}\BibitemShut
  {NoStop}%
\bibitem [{\citenamefont {Moriyama}(1995)}]{SolarAxion1995}%
  \BibitemOpen
\bibfield  {number} {  }\bibfield  {author} {\bibinfo {author} {\bibfnamefont
  {S.}~\bibnamefont {Moriyama}},\ }\href@noop {} {\bibfield  {journal}
  {\bibinfo  {journal} {Physical review letters}\ }\textbf {\bibinfo {volume}
  {75}},\ \bibinfo {pages} {3222} (\bibinfo {year} {1995})}\BibitemShut
  {NoStop}%
\bibitem [{\citenamefont {Dimopoulos}\ \emph {et~al.}(1986)\citenamefont
  {Dimopoulos}, \citenamefont {Frieman}, \citenamefont {Lynn},\ and\
  \citenamefont {Starkman}}]{SolarAxion1986}%
  \BibitemOpen
  \bibfield  {author} {\bibinfo {author} {\bibfnamefont {S.}~\bibnamefont
  {Dimopoulos}}, \bibinfo {author} {\bibfnamefont {J.}~\bibnamefont {Frieman}},
  \bibinfo {author} {\bibfnamefont {B.~W.}\ \bibnamefont {Lynn}},\ and\
  \bibinfo {author} {\bibfnamefont {G.~D.}\ \bibnamefont {Starkman}},\
  }\href@noop {} {\bibfield  {journal} {\bibinfo  {journal} {Physics Letters
  B}\ }\textbf {\bibinfo {volume} {179}},\ \bibinfo {pages} {223} (\bibinfo
  {year} {1986})}\BibitemShut {NoStop}%
\bibitem [{\citenamefont {Vilain}\ \emph {et~al.}(1993)\citenamefont {Vilain},
  \citenamefont {Wilquet}, \citenamefont {Beyer}, \citenamefont {Flegel},
  \citenamefont {Grote} \emph {et~al.}}]{veES_cross_section_measurement_1993}%
  \BibitemOpen
  \bibfield  {author} {\bibinfo {author} {\bibfnamefont {P.}~\bibnamefont
  {Vilain}}, \bibinfo {author} {\bibfnamefont {G.}~\bibnamefont {Wilquet}},
  \bibinfo {author} {\bibfnamefont {R.}~\bibnamefont {Beyer}}, \bibinfo
  {author} {\bibfnamefont {W.}~\bibnamefont {Flegel}}, \bibinfo {author}
  {\bibfnamefont {H.}~\bibnamefont {Grote}}, \emph {et~al.},\ }\href@noop {}
  {\bibfield  {journal} {\bibinfo  {journal} {Physics Letters B}\ }\textbf
  {\bibinfo {volume} {302}},\ \bibinfo {pages} {351} (\bibinfo {year}
  {1993})}\BibitemShut {NoStop}%
\bibitem [{\citenamefont {Auerbach}\ \emph {et~al.}(2001)\citenamefont
  {Auerbach}, \citenamefont {Burman}, \citenamefont {Caldwell}, \citenamefont
  {Church}, \citenamefont {Donahue} \emph
  {et~al.}}]{veES_cross_section_measurement_2001}%
  \BibitemOpen
  \bibfield  {author} {\bibinfo {author} {\bibfnamefont {L.}~\bibnamefont
  {Auerbach}}, \bibinfo {author} {\bibfnamefont {R.}~\bibnamefont {Burman}},
  \bibinfo {author} {\bibfnamefont {D.}~\bibnamefont {Caldwell}}, \bibinfo
  {author} {\bibfnamefont {E.}~\bibnamefont {Church}}, \bibinfo {author}
  {\bibfnamefont {J.}~\bibnamefont {Donahue}}, \emph {et~al.},\ }\href@noop {}
  {\bibfield  {journal} {\bibinfo  {journal} {Physical Review D}\ }\textbf
  {\bibinfo {volume} {63}},\ \bibinfo {pages} {112001} (\bibinfo {year}
  {2001})}\BibitemShut {NoStop}%
\bibitem [{\citenamefont {Bola{\~n}os}\ \emph {et~al.}(2009)\citenamefont
  {Bola{\~n}os}, \citenamefont {Miranda}, \citenamefont {Palazzo},
  \citenamefont {T{\'o}rtola},\ and\ \citenamefont
  {Valle}}]{veES_cross_section_theory_2009}%
  \BibitemOpen
  \bibfield  {author} {\bibinfo {author} {\bibfnamefont {A.}~\bibnamefont
  {Bola{\~n}os}}, \bibinfo {author} {\bibfnamefont {O.}~\bibnamefont
  {Miranda}}, \bibinfo {author} {\bibfnamefont {A.}~\bibnamefont {Palazzo}},
  \bibinfo {author} {\bibfnamefont {M.~A.}\ \bibnamefont {T{\'o}rtola}},\ and\
  \bibinfo {author} {\bibfnamefont {J.~W.}\ \bibnamefont {Valle}},\ }\href@noop
  {} {\bibfield  {journal} {\bibinfo  {journal} {Physical Review D}\ }\textbf
  {\bibinfo {volume} {79}},\ \bibinfo {pages} {113012} (\bibinfo {year}
  {2009})}\BibitemShut {NoStop}%
\bibitem [{\citenamefont {Tomalak}\ and\ \citenamefont
  {Hill}(2020)}]{veES_cross_section_theory_2020}%
  \BibitemOpen
  \bibfield  {author} {\bibinfo {author} {\bibfnamefont {O.}~\bibnamefont
  {Tomalak}}\ and\ \bibinfo {author} {\bibfnamefont {R.~J.}\ \bibnamefont
  {Hill}},\ }\href@noop {} {\bibfield  {journal} {\bibinfo  {journal} {Physical
  Review D}\ }\textbf {\bibinfo {volume} {101}},\ \bibinfo {pages} {033006}
  (\bibinfo {year} {2020})}\BibitemShut {NoStop}%
\bibitem [{\citenamefont {Vogel}\ and\ \citenamefont {Engel}(1989)}]{vMM_1989}%
  \BibitemOpen
  \bibfield  {author} {\bibinfo {author} {\bibfnamefont {P.}~\bibnamefont
  {Vogel}}\ and\ \bibinfo {author} {\bibfnamefont {J.}~\bibnamefont {Engel}},\
  }\href@noop {} {\bibfield  {journal} {\bibinfo  {journal} {Physical Review
  D}\ }\textbf {\bibinfo {volume} {39}},\ \bibinfo {pages} {3378} (\bibinfo
  {year} {1989})}\BibitemShut {NoStop}%
\bibitem [{\citenamefont {Wong}\ and\ \citenamefont {Li}(2005)}]{vMM_2005}%
  \BibitemOpen
  \bibfield  {author} {\bibinfo {author} {\bibfnamefont {H.~T.}\ \bibnamefont
  {Wong}}\ and\ \bibinfo {author} {\bibfnamefont {H.-B.}\ \bibnamefont {Li}},\
  }\href@noop {} {\bibfield  {journal} {\bibinfo  {journal} {Modern Physics
  Letters A}\ }\textbf {\bibinfo {volume} {20}},\ \bibinfo {pages} {1103}
  (\bibinfo {year} {2005})}\BibitemShut {NoStop}%
\bibitem [{\citenamefont {Giunti}\ and\ \citenamefont
  {Studenikin}(2009{\natexlab{b}})}]{vMM_review_2009}%
  \BibitemOpen
  \bibfield  {author} {\bibinfo {author} {\bibfnamefont {C.}~\bibnamefont
  {Giunti}}\ and\ \bibinfo {author} {\bibfnamefont {A.}~\bibnamefont
  {Studenikin}},\ }\href@noop {} {\bibfield  {journal} {\bibinfo  {journal}
  {Physics of Atomic Nuclei}\ }\textbf {\bibinfo {volume} {72}},\ \bibinfo
  {pages} {2089} (\bibinfo {year} {2009}{\natexlab{b}})}\BibitemShut {NoStop}%
\bibitem [{\citenamefont {Cleveland}\ \emph {et~al.}(1998)\citenamefont
  {Cleveland}, \citenamefont {Daily}, \citenamefont {Davis~Jr}, \citenamefont
  {Distel}, \citenamefont {Lande}, \citenamefont {Lee}, \citenamefont
  {Wildenhain},\ and\ \citenamefont {Ullman}}]{Homestake_solar_neutrinos_1998}%
  \BibitemOpen
  \bibfield  {author} {\bibinfo {author} {\bibfnamefont {B.~T.}\ \bibnamefont
  {Cleveland}}, \bibinfo {author} {\bibfnamefont {T.}~\bibnamefont {Daily}},
  \bibinfo {author} {\bibfnamefont {R.}~\bibnamefont {Davis~Jr}}, \bibinfo
  {author} {\bibfnamefont {J.~R.}\ \bibnamefont {Distel}}, \bibinfo {author}
  {\bibfnamefont {K.}~\bibnamefont {Lande}}, \bibinfo {author} {\bibfnamefont
  {C.}~\bibnamefont {Lee}}, \bibinfo {author} {\bibfnamefont {P.~S.}\
  \bibnamefont {Wildenhain}},\ and\ \bibinfo {author} {\bibfnamefont
  {J.}~\bibnamefont {Ullman}},\ }\href@noop {} {\bibfield  {journal} {\bibinfo
  {journal} {The Astrophysical Journal}\ }\textbf {\bibinfo {volume} {496}},\
  \bibinfo {pages} {505} (\bibinfo {year} {1998})}\BibitemShut {NoStop}%
\bibitem [{\citenamefont {Collaboration}\ \emph {et~al.}(2001)\citenamefont
  {Collaboration} \emph {et~al.}}]{SNO_solar_neutrinos_2001}%
  \BibitemOpen
  \bibfield  {author} {\bibinfo {author} {\bibfnamefont {S.}~\bibnamefont
  {Collaboration}} \emph {et~al.},\ }\href@noop {} {\bibfield  {journal}
  {\bibinfo  {journal} {Physical Review Letters}\ }\textbf {\bibinfo {volume}
  {87}},\ \bibinfo {pages} {71301} (\bibinfo {year} {2001})}\BibitemShut
  {NoStop}%
\bibitem [{\citenamefont {Fukuda}\ \emph {et~al.}(2001)\citenamefont {Fukuda},
  \citenamefont {Fukuda}, \citenamefont {Ishitsuka}, \citenamefont {Itow},
  \citenamefont {Kajita} \emph {et~al.}}]{superK_solar_neutrinos_2001}%
  \BibitemOpen
  \bibfield  {author} {\bibinfo {author} {\bibfnamefont {S.}~\bibnamefont
  {Fukuda}}, \bibinfo {author} {\bibfnamefont {Y.}~\bibnamefont {Fukuda}},
  \bibinfo {author} {\bibfnamefont {M.}~\bibnamefont {Ishitsuka}}, \bibinfo
  {author} {\bibfnamefont {Y.}~\bibnamefont {Itow}}, \bibinfo {author}
  {\bibfnamefont {T.}~\bibnamefont {Kajita}}, \emph {et~al.},\ }\href@noop {}
  {\bibfield  {journal} {\bibinfo  {journal} {Physical Review Letters}\
  }\textbf {\bibinfo {volume} {86}},\ \bibinfo {pages} {5656} (\bibinfo {year}
  {2001})}\BibitemShut {NoStop}%
\bibitem [{\citenamefont {Agostini}\ \emph {et~al.}(2018)\citenamefont
  {Agostini}, \citenamefont {Altenm{\"u}ller}, \citenamefont {Appel},
  \citenamefont {Jeschke}, \citenamefont {Neumair} \emph
  {et~al.}}]{Borexino_solar_pp_chain_neutrinos_2018}%
  \BibitemOpen
  \bibfield  {author} {\bibinfo {author} {\bibfnamefont {M.}~\bibnamefont
  {Agostini}}, \bibinfo {author} {\bibfnamefont {K.}~\bibnamefont
  {Altenm{\"u}ller}}, \bibinfo {author} {\bibfnamefont {S.}~\bibnamefont
  {Appel}}, \bibinfo {author} {\bibfnamefont {D.}~\bibnamefont {Jeschke}},
  \bibinfo {author} {\bibfnamefont {B.}~\bibnamefont {Neumair}}, \emph
  {et~al.},\ }\href@noop {} {\bibfield  {journal} {\bibinfo  {journal}
  {Nature}\ }\textbf {\bibinfo {volume} {562}},\ \bibinfo {pages} {505}
  (\bibinfo {year} {2018})}\BibitemShut {NoStop}%
\bibitem [{\citenamefont {Collaboration}\ \emph {et~al.}(2020)\citenamefont
  {Collaboration} \emph {et~al.}}]{Borexino_solar_CNO_cycle_neutrinos_2020}%
  \BibitemOpen
  \bibfield  {author} {\bibinfo {author} {\bibfnamefont {B.}~\bibnamefont
  {Collaboration}} \emph {et~al.},\ }\href@noop {} {\bibfield  {journal}
  {\bibinfo  {journal} {Nature}\ }\textbf {\bibinfo {volume} {587}},\ \bibinfo
  {pages} {577} (\bibinfo {year} {2020})}\BibitemShut {NoStop}%
\bibitem [{\citenamefont {Bahcall}\ \emph {et~al.}(1982)\citenamefont
  {Bahcall}, \citenamefont {Huebner}, \citenamefont {Lubow}, \citenamefont
  {Parker},\ and\ \citenamefont {Ulrich}}]{Standard_Solar_Model_1982}%
  \BibitemOpen
  \bibfield  {author} {\bibinfo {author} {\bibfnamefont {J.~N.}\ \bibnamefont
  {Bahcall}}, \bibinfo {author} {\bibfnamefont {W.~F.}\ \bibnamefont
  {Huebner}}, \bibinfo {author} {\bibfnamefont {S.~H.}\ \bibnamefont {Lubow}},
  \bibinfo {author} {\bibfnamefont {P.~D.}\ \bibnamefont {Parker}},\ and\
  \bibinfo {author} {\bibfnamefont {R.~K.}\ \bibnamefont {Ulrich}},\
  }\href@noop {} {\bibfield  {journal} {\bibinfo  {journal} {Reviews of Modern
  Physics}\ }\textbf {\bibinfo {volume} {54}},\ \bibinfo {pages} {767}
  (\bibinfo {year} {1982})}\BibitemShut {NoStop}%
\bibitem [{\citenamefont {Bahcall}\ and\ \citenamefont
  {Pinsonneault}(1992)}]{Standard_Solar_Model_1992}%
  \BibitemOpen
  \bibfield  {author} {\bibinfo {author} {\bibfnamefont {J.~N.}\ \bibnamefont
  {Bahcall}}\ and\ \bibinfo {author} {\bibfnamefont {M.}~\bibnamefont
  {Pinsonneault}},\ }\href@noop {} {\bibfield  {journal} {\bibinfo  {journal}
  {Reviews of Modern Physics}\ }\textbf {\bibinfo {volume} {64}},\ \bibinfo
  {pages} {885} (\bibinfo {year} {1992})}\BibitemShut {NoStop}%
\bibitem [{\citenamefont {Bahcall}\ \emph {et~al.}(2006)\citenamefont
  {Bahcall}, \citenamefont {Serenelli},\ and\ \citenamefont
  {Basu}}]{Standard_Solar_Model_2006}%
  \BibitemOpen
  \bibfield  {author} {\bibinfo {author} {\bibfnamefont {J.~N.}\ \bibnamefont
  {Bahcall}}, \bibinfo {author} {\bibfnamefont {A.~M.}\ \bibnamefont
  {Serenelli}},\ and\ \bibinfo {author} {\bibfnamefont {S.}~\bibnamefont
  {Basu}},\ }\href@noop {} {\bibfield  {journal} {\bibinfo  {journal} {The
  Astrophysical Journal Supplement Series}\ }\textbf {\bibinfo {volume}
  {165}},\ \bibinfo {pages} {400} (\bibinfo {year} {2006})}\BibitemShut
  {NoStop}%
\bibitem [{\citenamefont {Vinyoles}\ \emph {et~al.}(2017)\citenamefont
  {Vinyoles}, \citenamefont {Serenelli}, \citenamefont {Villante},
  \citenamefont {Basu}, \citenamefont {Bergstr{\"o}m}, \citenamefont
  {Gonzalez-Garcia}, \citenamefont {Maltoni}, \citenamefont {Pe{\~n}a-Garay},\
  and\ \citenamefont {Song}}]{new_Standard_Solar_Model_2017}%
  \BibitemOpen
  \bibfield  {author} {\bibinfo {author} {\bibfnamefont {N.}~\bibnamefont
  {Vinyoles}}, \bibinfo {author} {\bibfnamefont {A.~M.}\ \bibnamefont
  {Serenelli}}, \bibinfo {author} {\bibfnamefont {F.~L.}\ \bibnamefont
  {Villante}}, \bibinfo {author} {\bibfnamefont {S.}~\bibnamefont {Basu}},
  \bibinfo {author} {\bibfnamefont {J.}~\bibnamefont {Bergstr{\"o}m}}, \bibinfo
  {author} {\bibfnamefont {M.}~\bibnamefont {Gonzalez-Garcia}}, \bibinfo
  {author} {\bibfnamefont {M.}~\bibnamefont {Maltoni}}, \bibinfo {author}
  {\bibfnamefont {C.}~\bibnamefont {Pe{\~n}a-Garay}},\ and\ \bibinfo {author}
  {\bibfnamefont {N.}~\bibnamefont {Song}},\ }\href@noop {} {\bibfield
  {journal} {\bibinfo  {journal} {The Astrophysical Journal}\ }\textbf
  {\bibinfo {volume} {835}},\ \bibinfo {pages} {202} (\bibinfo {year}
  {2017})}\BibitemShut {NoStop}%
\bibitem [{Joh()}]{John_Bahcall_website_1}%
  \BibitemOpen
  \href@noop {} {\bibinfo {title} {{John Bahcall's homepage:}}},\ \bibinfo
  {howpublished} {\url{http://www.sns.ias.edu/~jnb/}},\ \bibinfo {note}
  {accessed: 2020-12-20}\BibitemShut {NoStop}%
\bibitem [{\citenamefont {Alimonti}\ \emph {et~al.}(1998)\citenamefont
  {Alimonti}, \citenamefont {Angloher}, \citenamefont {Arpesella},
  \citenamefont {Balata}, \citenamefont {Bellini} \emph
  {et~al.}}]{Borexino_C14_1998}%
  \BibitemOpen
  \bibfield  {author} {\bibinfo {author} {\bibfnamefont {G.}~\bibnamefont
  {Alimonti}}, \bibinfo {author} {\bibfnamefont {G.}~\bibnamefont {Angloher}},
  \bibinfo {author} {\bibfnamefont {C.}~\bibnamefont {Arpesella}}, \bibinfo
  {author} {\bibfnamefont {M.}~\bibnamefont {Balata}}, \bibinfo {author}
  {\bibfnamefont {G.}~\bibnamefont {Bellini}}, \emph {et~al.},\ }\href@noop {}
  {\bibfield  {journal} {\bibinfo  {journal} {Physics letters B}\ }\textbf
  {\bibinfo {volume} {422}},\ \bibinfo {pages} {349} (\bibinfo {year}
  {1998})}\BibitemShut {NoStop}%
\bibitem [{\citenamefont {Bellini}\ \emph {et~al.}(2014)\citenamefont
  {Bellini}, \citenamefont {Benziger}, \citenamefont {Bick}, \citenamefont
  {Bonfini}, \citenamefont {Bravo} \emph
  {et~al.}}]{Borexino_solar_pp_chain_C14_2014}%
  \BibitemOpen
  \bibfield  {author} {\bibinfo {author} {\bibfnamefont {G.}~\bibnamefont
  {Bellini}}, \bibinfo {author} {\bibfnamefont {J.}~\bibnamefont {Benziger}},
  \bibinfo {author} {\bibfnamefont {D.}~\bibnamefont {Bick}}, \bibinfo {author}
  {\bibfnamefont {G.}~\bibnamefont {Bonfini}}, \bibinfo {author} {\bibfnamefont
  {D.}~\bibnamefont {Bravo}}, \emph {et~al.},\ }\href@noop {} {\bibfield
  {journal} {\bibinfo  {journal} {Nature}\ }\textbf {\bibinfo {volume} {512}},\
  \bibinfo {pages} {383} (\bibinfo {year} {2014})}\BibitemShut {NoStop}%
\bibitem [{\citenamefont {Li}\ \emph {et~al.}(2011)\citenamefont {Li},
  \citenamefont {Xiao}, \citenamefont {Cao}, \citenamefont {Li}, \citenamefont
  {Ruan},\ and\ \citenamefont {Heng}}]{PSD_liquid_scintillator}%
  \BibitemOpen
  \bibfield  {author} {\bibinfo {author} {\bibfnamefont {X.-B.}\ \bibnamefont
  {Li}}, \bibinfo {author} {\bibfnamefont {H.-L.}\ \bibnamefont {Xiao}},
  \bibinfo {author} {\bibfnamefont {J.}~\bibnamefont {Cao}}, \bibinfo {author}
  {\bibfnamefont {J.}~\bibnamefont {Li}}, \bibinfo {author} {\bibfnamefont
  {X.-C.}\ \bibnamefont {Ruan}},\ and\ \bibinfo {author} {\bibfnamefont
  {Y.-K.}\ \bibnamefont {Heng}},\ }\href@noop {} {\bibfield  {journal}
  {\bibinfo  {journal} {Chinese Physics C}\ }\textbf {\bibinfo {volume} {35}},\
  \bibinfo {pages} {1026} (\bibinfo {year} {2011})}\BibitemShut {NoStop}%
\bibitem [{\citenamefont {Abusleme}\ \emph {et~al.}(2020)\citenamefont
  {Abusleme}, \citenamefont {Adam}, \citenamefont {Ahmad}, \citenamefont
  {Aiello}, \citenamefont {Akram} \emph
  {et~al.}}]{JUNO_solar_B8_neutrinos_2020}%
  \BibitemOpen
  \bibfield  {author} {\bibinfo {author} {\bibfnamefont {A.}~\bibnamefont
  {Abusleme}}, \bibinfo {author} {\bibfnamefont {T.}~\bibnamefont {Adam}},
  \bibinfo {author} {\bibfnamefont {S.}~\bibnamefont {Ahmad}}, \bibinfo
  {author} {\bibfnamefont {S.}~\bibnamefont {Aiello}}, \bibinfo {author}
  {\bibfnamefont {M.}~\bibnamefont {Akram}}, \emph {et~al.},\ }\href@noop {}
  {\bibfield  {journal} {\bibinfo  {journal} {arXiv preprint arXiv:2006.11760}\
  } (\bibinfo {year} {2020})}\BibitemShut {NoStop}%
\bibitem [{\citenamefont {Zhang}\ \emph {et~al.}(2018)\citenamefont {Zhang},
  \citenamefont {He}, \citenamefont {Li},\ and\ \citenamefont
  {Xu}}]{muon_tracking_JUNO_2018}%
  \BibitemOpen
  \bibfield  {author} {\bibinfo {author} {\bibfnamefont {K.}~\bibnamefont
  {Zhang}}, \bibinfo {author} {\bibfnamefont {M.}~\bibnamefont {He}}, \bibinfo
  {author} {\bibfnamefont {W.}~\bibnamefont {Li}},\ and\ \bibinfo {author}
  {\bibfnamefont {J.}~\bibnamefont {Xu}},\ }\href@noop {} {\bibfield  {journal}
  {\bibinfo  {journal} {Radiation Detection Technology and Methods}\ }\textbf
  {\bibinfo {volume} {2}},\ \bibinfo {pages} {13} (\bibinfo {year}
  {2018})}\BibitemShut {NoStop}%
\bibitem [{\citenamefont {Abusleme}\ \emph {et~al.}(2021)\citenamefont
  {Abusleme}, \citenamefont {Adam}, \citenamefont {Ahmad}, \citenamefont
  {Ahmed}, \citenamefont {Aiello} \emph {et~al.}}]{JUNO_calibration_2020}%
  \BibitemOpen
  \bibfield  {author} {\bibinfo {author} {\bibfnamefont {A.}~\bibnamefont
  {Abusleme}}, \bibinfo {author} {\bibfnamefont {T.}~\bibnamefont {Adam}},
  \bibinfo {author} {\bibfnamefont {S.}~\bibnamefont {Ahmad}}, \bibinfo
  {author} {\bibfnamefont {R.}~\bibnamefont {Ahmed}}, \bibinfo {author}
  {\bibfnamefont {S.}~\bibnamefont {Aiello}}, \emph {et~al.},\ }\href@noop {}
  {\bibfield  {journal} {\bibinfo  {journal} {Journal of High Energy Physics}\
  }\textbf {\bibinfo {volume} {2021}},\ \bibinfo {pages} {1} (\bibinfo {year}
  {2021})}\BibitemShut {NoStop}%
\bibitem [{JUN()}]{JUNO_20_inch_PMT}%
  \BibitemOpen
  \href@noop {} {\bibinfo {title} {{NEUTRINO2020 poster:}}},\ \bibinfo
  {howpublished}
  {\url{https://nusoft.fnal.gov/nova/nu2020postersession/pdf/posterPDF-370.pdf}},\
  \bibinfo {note} {accessed: 2021-1-19}\BibitemShut {NoStop}%
\bibitem [{\citenamefont {Cochran}(1952)}]{chi_square_fit_1952}%
  \BibitemOpen
  \bibfield  {author} {\bibinfo {author} {\bibfnamefont {W.~G.}\ \bibnamefont
  {Cochran}},\ }\href@noop {} {\bibfield  {journal} {\bibinfo  {journal} {The
  Annals of mathematical statistics}\ }\textbf {\bibinfo {volume} {23}},\
  \bibinfo {pages} {315} (\bibinfo {year} {1952})}\BibitemShut {NoStop}%
\bibitem [{MM_()}]{MM_trigger_1}%
  \BibitemOpen
  \href@noop {} {\bibinfo {title} {{NEUTRINO2020 poster:}}},\ \bibinfo
  {howpublished}
  {\url{https://nusoft.fnal.gov/nova/nu2020postersession/pdf/posterPDF-129.pdf}},\
  \bibinfo {note} {accessed: 2021-1-19}\BibitemShut {NoStop}%
\bibitem [{\citenamefont {Andringa}\ \emph {et~al.}(2016)\citenamefont
  {Andringa} \emph {et~al.}}]{SNO_2016}%
  \BibitemOpen
  \bibfield  {author} {\bibinfo {author} {\bibfnamefont {S.}~\bibnamefont
  {Andringa}} \emph {et~al.} (\bibinfo {collaboration} {SNO+}),\ }\href
  {https://doi.org/10.1155/2016/6194250} {\bibfield  {journal} {\bibinfo
  {journal} {Adv. High Energy Phys.}\ }\textbf {\bibinfo {volume} {2016}},\
  \bibinfo {pages} {6194250} (\bibinfo {year} {2016})},\ \Eprint
  {https://arxiv.org/abs/1508.05759} {arXiv:1508.05759 [physics.ins-det]}
  \BibitemShut {NoStop}%
\bibitem [{\citenamefont {Kastens}\ \emph {et~al.}(2009)\citenamefont
  {Kastens}, \citenamefont {Cahn}, \citenamefont {Manzur},\ and\ \citenamefont
  {McKinsey}}]{Kastens_2009}%
  \BibitemOpen
  \bibfield  {author} {\bibinfo {author} {\bibfnamefont {L.~W.}\ \bibnamefont
  {Kastens}}, \bibinfo {author} {\bibfnamefont {S.~B.}\ \bibnamefont {Cahn}},
  \bibinfo {author} {\bibfnamefont {A.}~\bibnamefont {Manzur}},\ and\ \bibinfo
  {author} {\bibfnamefont {D.~N.}\ \bibnamefont {McKinsey}},\ }\bibfield
  {journal} {\bibinfo  {journal} {Physical Review C}\ }\textbf {\bibinfo
  {volume} {80}},\ \href {https://doi.org/10.1103/physrevc.80.045809}
  {10.1103/physrevc.80.045809} (\bibinfo {year} {2009})\BibitemShut {NoStop}%
\bibitem [{\citenamefont {Manalaysay}\ \emph {et~al.}(2010)\citenamefont
  {Manalaysay}, \citenamefont {Undagoitia}, \citenamefont {Askin},
  \citenamefont {Baudis}, \citenamefont {Behrens}, \citenamefont {Ferella},
  \citenamefont {Kish}, \citenamefont {Lebeda}, \citenamefont {Santorelli},
  \citenamefont {Vénos},\ and\ \citenamefont {et~al.}}]{Manalaysay_2010}%
  \BibitemOpen
  \bibfield  {author} {\bibinfo {author} {\bibfnamefont {A.}~\bibnamefont
  {Manalaysay}}, \bibinfo {author} {\bibfnamefont {T.~M.}\ \bibnamefont
  {Undagoitia}}, \bibinfo {author} {\bibfnamefont {A.}~\bibnamefont {Askin}},
  \bibinfo {author} {\bibfnamefont {L.}~\bibnamefont {Baudis}}, \bibinfo
  {author} {\bibfnamefont {A.}~\bibnamefont {Behrens}}, \bibinfo {author}
  {\bibfnamefont {A.~D.}\ \bibnamefont {Ferella}}, \bibinfo {author}
  {\bibfnamefont {A.}~\bibnamefont {Kish}}, \bibinfo {author} {\bibfnamefont
  {O.}~\bibnamefont {Lebeda}}, \bibinfo {author} {\bibfnamefont
  {R.}~\bibnamefont {Santorelli}}, \bibinfo {author} {\bibfnamefont
  {D.}~\bibnamefont {Vénos}},\ and\ \bibinfo {author} {\bibnamefont
  {et~al.}},\ }\href {https://doi.org/10.1063/1.3436636} {\bibfield  {journal}
  {\bibinfo  {journal} {Review of Scientific Instruments}\ }\textbf {\bibinfo
  {volume} {81}},\ \bibinfo {pages} {073303} (\bibinfo {year}
  {2010})}\BibitemShut {NoStop}%
\end{thebibliography}%
\bibliographystyle{apsrev4-2}

\end{document}